\documentclass[12pt]{article}
\usepackage{breakcites}
\usepackage[round,sort,comma,authoryear]{natbib} 
\usepackage[margin=1.5in]{geometry}
\usepackage[font=footnotesize]{caption}

\usepackage{setspace} 
\usepackage[colorlinks=true,linkcolor=magenta,citecolor=magenta,backref=page]{hyperref}

\usepackage{enumerate}
\usepackage{amsmath}
\usepackage{amssymb}
\usepackage{amsthm}
\usepackage{mathtools}
\usepackage{bm}

\usepackage{graphicx}
\usepackage{float}
\usepackage{tikz}
\usepackage{subfig}

\newtheorem{proposition}{Proposition}
\newtheorem{theorem}{Theorem}

\newtheorem{corollary}{Corollary}

{
\newtheorem{definition}{Definition}}

\title{Surplus Extraction with Behavioral Types\thanks{I thank Luca Rigotti and Richard Van Weelden for providing guidance in this project. I also thank Svetlana Kosterina, Alexey Kushnir, Benjamin Matta, Tymofiy Mylovanov, participants of the Microeconomic Theory Brownbag at University of Pittsburgh, the 2021 Pennsylvania Economic Theory Conference and the 32nd Stony Brook International Conference on Game Theory for useful comments.}}
	
\author{Nicolas Pastrian\thanks{Department of Economics, University of Pittsburgh. Email: nip59@pitt.edu}}
\date{This version: \today\\\href{https://nrpastrian.github.io/files/surplus_extraction_behavioral.pdf}{For the most recent version click here}}

\begin{document}
\maketitle
	\begin{abstract}
		We examine the surplus extraction problem in a mechanism design setting with behavioral types. In our model behavioral types always perfectly reveal their private information.  We characterize the sufficient conditions that guarantee full extraction in a finite version of the reduced form environment of \citet{mcafeereny1992}. We found that the standard convex independence condition identified in \citet{cremermclean1988} is required only among the beliefs of strategic types, while a weaker condition is required for the beliefs of behavioral types. 
	\end{abstract}

\section{Introduction}
In mechanism design settings, private information leads agents to retain informational rents if information is independently distributed. However, we have known since \citet{myerson1981} that under correlation it is possible to extract all the informational rents from agents. Hence, in the presence of correlation, private information does not necessarily lead to agents obtaining information rents. This result is what is usually called \textit{full (surplus) extraction}. \citet{cremermclean1988} have identified the key independence condition that guarantees full extraction (convex independence) in mechanism design settings. Moreover, this condition remains key to guarantee surplus extraction in more general environments (see for example \citet{farinhaluz2013}, \citet{krahmer2020}, and \citet{fuetal2021samples}).

We examine the full surplus extraction problem in the presence of \textit{behavioral types}. That is, types that don't respond optimally to incentives. We focus on a particular case of such behavioral types: types that don't react to the mechanisms implemented and always reveal their private information perfectly. This assumption simplifies the characterization of the model and works as a benchmark for the designer's problem since it gives him the most advantaged setting to operate. The key feature of this assumption is that it allows for a perfect identification of each behavioral type and allows us to characterize their behavior regardless of the mechanism implemented. While extracting rents from behavioral types should be easier, incentive compatibility still requires their contract to be non-attractive to strategic types, imposing constraints on the designer. 

We consider a reduced form environment similar to the one used by \citet{mcafeereny1992} and \citet{lopomorigottishannon_detectability}, where in an unmodeled stage an agent is left with informational rents that depend on his private information. The contracts can condition on an exogenous source of uncertainty. We will refer to the current stage contracts simply as contracts and ignore any reference to the mechanism that generates the informational rents. The main differences with respect the model in \citet{mcafeereny1992} are that we introduce a class of behavioral types which always report their private information truthfully and focus on a finite type space. 

Behavioral types that report their information truthfully have been studied before by \citet{severinov_deneckere2006} and \citet{saran2011} in the context of monopolistic screening and bilateral trade respectively. However, correlation plays no role in their models. More general models from behavioral economics have been studied in the context of implementation (see for example \citet{eliaz2002} and \citet{declippeletal2018}).

Our main result is a characterization of the conditions that guarantee full extraction in a environment with correlation and behavioral types. The key condition is a relaxation of the convex independence condition identified by \citet{cremermclean1988} for the beliefs of the behavioral types. This condition is not only required to guarantee full extraction over all types (Theorem \ref{thm}) but also allows extraction of all the surplus from behavioral types even if the full surplus extraction result fails (Corollary \ref{corollary}). 

The remainder of this note is organized as follows. We describe the model and results in Section \ref{sec:model}. In Section \ref{sec:auction} we apply our main result to an auction environment with correlated valuations. Finally, Section \ref{sec:end} concludes.

\section{Model}\label{sec:model}

There is a finite set of types $T$ and a finite set of states $\Omega$. Each type $t$ is associated with a valuation (i.e.,  informational rents) $v_t\in\mathbb{R}_+$ and beliefs $p_t\in\Delta(\Omega)$.\footnote{In Section \ref{sec:auction} we explicitly derive $v_t$ and $p_t$ in an auction example.} We define a contract $c$ as a mapping from states into transfers, such that $c(\omega)\in \mathbb{R}$ is the transfer required in state $\omega\in\Omega$. A contract menu $\mathbf{c}$ is a collection $\{c_t: t\in T\}$ such that $c_t:\Omega \to \mathbb{R}$, i.e., a collection of contracts, one for each type.

There is a single agent with quasilinear preferences. Hence, from a contract $c$, type $t$'s payoff is given by
	\[
		v_t-\left<p_t,c\right>
	\]
where $\left<p_t,c\right>$ denotes the expected value of $c$ under $p_t$, that is 
	\[
		\left<p_t,c\right>=\sum_{\omega\in \Omega} p_t(\omega)\,c(\omega).
	\]

We are interested in whether the principal or designer is able to extract all the rents from the agent using a contract menu $\mathbf{c}$.\footnote{However, as \citet{borgersbook} notes, the focus on surplus extraction is arbitrary and the same results could be applied to implement any particular profile of payoffs or even allocations.}


We introduce the definition of full extraction formally.

\begin{definition}\label{def_fse}
	A contract menu $\mathbf{c}$ achieves full extraction if for all $t\in T$
		\[
			\left<p_t,c_t\right>=v_t.
		\]
\end{definition}

In the traditional setting it is required that all types prefer their own contract to the contract offered to others. We depart from the standard model by introducing a particular class of \emph{behavioral} types. In particular, a behavioral type is a type for which no incentive compatibility constraint is considered. That is, a behavioral type always reports his type truthfully. While this assumption could seem extreme, we claim that this is without loss due to a revelation principle argument as long as the strategy of the behavioral types is independent of the mechanism implemented. 

Let $B\subseteq T$ be the set of behavioral or unsophisticated types. Similarly, let $S=T\backslash B$ be the set of \emph{standard, sophisticated or strategic} types. 

Since in our model behavioral types don't respond to incentives, we will require incentive compatible constraints only for strategic types. Moreover, we will use a restrictive incentive compatibility notion, requiring that potential deviations have no impact in the informational rents of the agent. That is, we will require that each (strategic) type chooses his cost minimizing contract. 

	\begin{definition}\label{def:ic}
		A contract menu $\mathbf{c}$ is incentive compatible if for each strategic type $s\in S$,
			\[
				c_s\in\underset{t\in T}{\arg\min} \left<p_s,c_t\right>.
			\]

	\end{definition}

We will look for an incentive compatible contract menu that fully extracts the informational rents from the agent. 

	\begin{definition}
		Full extraction with behavioral types is feasible if there exists an incentive compatible contract menu $\mathbf{c}$ which achieves full extraction.
	\end{definition}
\citet{cremermclean1988} have shown that in a setting without behavioral types full extraction is feasible if the set of beliefs satisfies the independence condition in Definition \ref{def:cm}.

\begin{definition}\label{def:cm}
	A set of beliefs $P$ satisfies the CM condition if for any $p\in P$, $p\not\in co(P\backslash\{p\})$.
\end{definition}

This condition is known as the convex independence condition, and it is a linear independence condition over the set of beliefs. It also coincides with the more general condition of probabilistic independence used by \citet{mcafeereny1992} and \citet{lopomorigottishannon_detectability} if applied to a setting with finite types as the model we use here.

We assume that different types hold different beliefs, that is, $p_t\neq p_{t'}$ if $t \neq t'$, and denote by $P^X$ the set of beliefs associated to types in $X\subseteq T$. Note that $p_t\neq p_{t'}$ introduces correlation in our environment, i.e., types and states are not independent.

We proceed to present our main result.

\begin{theorem}\label{thm}
	Full extraction with behavioral types is feasible if
		\begin{enumerate}[(i)]
			\item $P^S$ satisfies the CM condition, and
			\item For all types $b\in B$, $p_b\not\in co(P^S)$.
		\end{enumerate}
\end{theorem}

\begin{proof}
Step 1: from (i), we know from \citet{cremermclean1988} that we can find a contract menu that reaches full extraction if we restrict types to $S$. Notice that such contract menu remains incentive compatible and reaches full extraction among types in $S$ as long as the contracts offered to types in $B$ don't generate incentives to any type in $S$ to deviate. 

Step 2: now, we construct the contracts for types in $B$. Consider a single behavioral type $b$. For this type, it suffices to find $z_b$ such that 
	\[
		\left<p_b,z_b\right>=0
	\]
	\[
		\left<p_s,z_b\right> > 0,\quad \forall s\in S 	
	\]
Due to condition (ii), the existence of a solution to this system of inequalities follows from Farkas' lemma. Hence, we can use $z_b$ to construct the extracting contract for the behavioral type in a similar way we construct the contract for types in $S$. In particular,
	\[c_b=v_b+\alpha_bz_b\]
with $\alpha_b=\max_{s\in S} \frac{v_s-v_b}{\left<p_s,z_b\right>}$ satisfies all the required conditions. 

Note that the contract designed for type $b$ has no impact on the contract of any other type in $B$ or $S$. Hence, we can repeat the process for all others types in $B$ to construct the contract for each remaining type.

\end{proof}

Notice that from the proof of Theorem \ref{thm}, it is without loss to look at environments where $|B|=1$ since the problem for each behavioral type could be looked at in complete isolation from other behavioral types. This is possible since there are no ``cross'' incentive compatibility conditions. 

Theorem \ref{thm} shows that a slightly relaxed convex independence condition is required to guarantee full extraction in the presence of behavioral types. Even though we have completely relaxed the incentive compatibility constraints for behavioral types, identification conditions in terms of their beliefs are still required to be able to extract all their rents. This highlights that incentive compatibility and surplus extraction are not isolated features of a mechanism. 

The second condition of the theorem allows us to separate a behavioral type from the strategic types. Moreover, this separation translates into a particular direction of incentive compatibility: that no strategic type wants to deviate to the contract of that particular behavioral type considered. This is in contrast to the first condition which requires, as in the standard extraction problem, to consider incentive compatibility in both directions. This is why the augmented condition required for behavioral types is weaker than the one required over strategic types. This also shows why the first condition cannot be relaxed, i.e., that $P^S$ must satisfy the standard CM condition: introducing behavioral types enlarges the set of constraints the designer must satisfy relative to the setting in which only types in $S$ are present, so separation among types in $S$ is still required. 




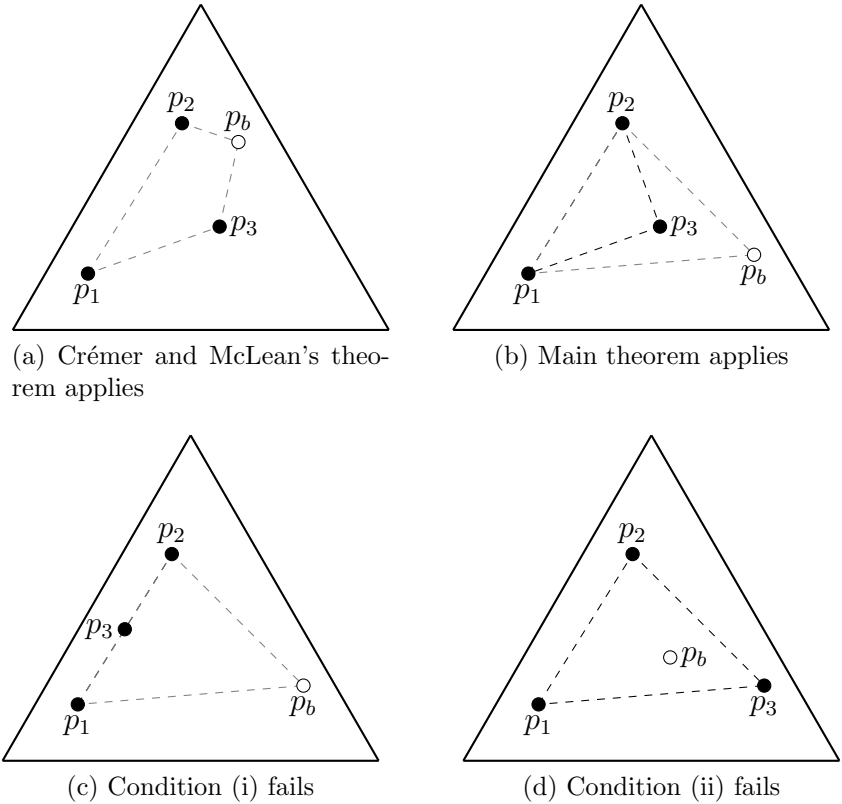
\begin{figure}[hb!]
\centering
\subfloat[Cr\'emer and McLean's theorem applies]{\begin{tikzpicture}[scale=2.5] 
	\draw[gray,dashed](.4,.3)to(.9,1.1)to(1.2,1)to(1.1,.55)to(.4,.3);

	\draw[black, fill] (.4,.3) circle (1pt); 
	\draw[black,below] (.4,.3) node{$p_1$}; 


	\draw[black, fill] (1.1,.55) circle (1pt); 
	\draw[black,right] (1.1,.55) node{$p_3$}; 

	\draw[black, fill] (.9,1.1) circle (1pt); 
	\draw[black,above] (.9,1.1) node{$p_2$}; 

	\draw[black,fill=white] (1.2,1) circle (1pt);
	\draw[black,above] (1.2,1) node{$p_b$};

	\draw [thick,domain=0:1] plot (\x, {sqrt(3)*\x});
	\draw [thick,domain=1:2] plot (\x, {2*sqrt(3)-sqrt(3)*\x});
	\draw[thick](0,0)to(2,0);

\end{tikzpicture}}\qquad
\subfloat[Main theorem applies]{\begin{tikzpicture}[scale=2.5]
	\draw[black,dashed](.4,.3)to(.9,1.1)to(1.1,.55)to(.4,.3);
	\draw[gray,dashed](.4,.3)to(.9,1.1)to(1.6,.4)to(.4,.3);
	\draw[black, fill] (.4,.3) circle (1pt); 
	\draw[black,below] (.4,.3) node{$p_1$};

	\draw[black, fill=white] (1.6,.4) circle (1pt);
	\draw[black,below] (1.6,.4) node{$p_b$};

	\draw[black, fill] (.9,1.1) circle (1pt);
	\draw[black,above] (.9,1.1) node{$p_2$};

	\draw[black, fill] (1.1,.55) circle (1pt);
	\draw[black,right] (1.1,.55) node{$p_3$};

	\draw [thick,domain=0:1] plot (\x, {sqrt(3)*\x});
	\draw [thick,domain=1:2] plot (\x, {2*sqrt(3)-sqrt(3)*\x});
	\draw[thick](0,0)to(2,0);

\end{tikzpicture}}\\
\subfloat[Condition (i) fails]{
	\begin{tikzpicture}[scale=2.5] 
	\draw[black,dashed](.4,.3)to(.9,1.1);
	\draw[gray,dashed](.4,.3)to(.9,1.1)to(1.6,.4)to(.4,.3);

	\draw[black, fill] (.4,.3) circle (1pt);
	\draw[black,below] (.4,.3) node{$p_1$};

	\draw[black, fill=white] (1.6,.4) circle (1pt);
	\draw[black,below] (1.6,.4) node{$p_b$};

	\draw[black, fill] (.9,1.1) circle (1pt);
	\draw[black,above] (.9,1.1) node{$p_2$};

	\draw[black, fill] (.65,.7) circle (1pt);
	\draw[black,left] (.65,.7) node{$p_3$};

	\draw [thick,domain=0:1] plot (\x, {sqrt(3)*\x});
	\draw [thick,domain=1:2] plot (\x, {2*sqrt(3)-sqrt(3)*\x});
	\draw[thick](0,0)to(2,0);

\end{tikzpicture}
}\qquad
\subfloat[Condition (ii) fails]{
	\begin{tikzpicture}[scale=2.5] 
		\draw[black,dashed](.4,.3)to(.9,1.1)to(1.6,.4)to(.4,.3);

		\draw[black, fill] (.4,.3) circle (1pt);
		\draw[black,below] (.4,.3) node{$p_1$};

		\draw[black, fill] (1.6,.4) circle (1pt);
		\draw[black,below] (1.6,.4) node{$p_3$};

		\draw[black, fill] (.9,1.1) circle (1pt);
		\draw[black,above] (.9,1.1) node{$p_2$};

		\draw[black, fill=white] (1.1,.55) circle (1pt);
		\draw[black,right] (1.1,.55) node{$p_b$};

		\draw [thick,domain=0:1] plot (\x, {sqrt(3)*\x});
		\draw [thick,domain=1:2] plot (\x, {2*sqrt(3)-sqrt(3)*\x});
		\draw[thick](0,0)to(2,0);

	\end{tikzpicture}
}
	\caption{Representation of Theorem \ref{thm} with three states.}

	\label{fig1}
\end{figure}

Figure \ref{fig1} illustrates the statements presented in Theorem \ref{thm} using a three states and four types with a single behavioral type example. In panel $a$, all beliefs are linearly independent, hence $P^T$ satisfies the CM condition and the original theorem from \citet{cremermclean1988} guarantees that full extraction is feasible. Instead in panel $b$, it is Theorem \ref{thm} that guarantees full surplus extraction. Clearly, if we consider all beliefs depicted there the standard CM condition is violated. However, the beliefs of strategic types in panel b ($p_1,p_2$ and $p_3$) does satisfy the CM condition (condition (i) in Theorem \ref{thm}), and the beliefs of the behavioral type ($p_b$) could not be represented as a convex combination of the beliefs of the strategic types (condition (ii) in Theorem \ref{thm}), hence by Theorem \ref{thm} we know full extraction is feasible in this case. Finally, panel $c$ and $d$ illustrate how the violation of each condition in Theorem \ref{thm} would looks like in the space of beliefs.


We finish this section presenting two results regarding violations of the conditions identified in Theorem \ref{thm}.

The first result is a direct consequence of Theorem \ref{thm}, and shows that the same contract works for behavioral types even if full extraction could not be achieved for strategic types.

\begin{corollary}\label{corollary}
	Consider a particular behavioral type $b\in B$. Let $c_{-b}$ be an incentive compatible contract menu for types $t\neq b$. If $p_b\not\in co(P^S)$ then there exists a contract $c_b$ such that the contract menu $(c_b,c_{-b})$ is incentive compatible and $\left<p_b,c_b\right>=v_b$.

\end{corollary}

	\begin{proof}
		Follows directly from the proof of Theorem \ref{thm}.
	\end{proof}

This result is analogous to the result in \citet{borgersbook} which shows that if the convex independence condition holds then any allocation rule could be made incentive compatible. Here we show that implementing full extraction for a particular behavioral type $b$ requires us to check only if his beliefs are in the convex hull of the beliefs of the strategic types, regardless of what could be achieved from other types.

Proposition \ref{prop} identifies a sufficient condition to guarantee full extraction even if the second condition in Theorem \ref{thm} fails. It involves imposing restrictions over the valuations of behavioral types with beliefs in the convex hull of the beliefs of strategic types.

\begin{proposition}\label{prop}
	Suppose $P^S$ satisfies the CM condition. Let $\hat{B}=\{b\in B: p_b\in co(P^S)\}$. Then, full extraction with behavioral types is feasible if for each $b\in\hat{B}$ 
	    \[
	     v_b\geq \sum_{s\in S} \lambda^b(s) v_s,
	    \]
	where $\lambda^b\in\Delta(S): p_b=\displaystyle\sum_{s\in S} \lambda^b(s) p_s$.
\end{proposition}

\begin{proof}
	The contracts for types $s\in S$ and $b\in B\backslash \hat{B}$ remain the same as in Theorem \ref{thm}.

	For types $b\in \hat{B}$, there are two cases to consider:

		\begin{enumerate}
			\item If $v_b\geq \max_{s\in S}v_s$ then a flat constant contract $c_b=v_b$ doesn't violate any incentive compatibility condition and extract all rents from type $b$. 

			\item For $v_b<\max_{s\in S} v_s$ we construct the contract for $b$ as follows. 

			First, $p_b\in co(P^S)$ implies there exist weights $(\lambda_s)_{s\in S}$ such that $\sum_{s\in S}
			\lambda_s=1$, $\lambda_s\geq 0$ for all $s\in S$ and $p_b=\sum_{s\in S} \lambda_sp_s$.

			Let $v_{max}=\max_{s\in S} v_s$ and $\overline{v}_b=\sum_{s\in S} \lambda_s v_s$.  Now, consider the contract

				\[
					c_b=\alpha_b\sum_{s_\in S} \lambda_s c_s +(1-\alpha_b) v_{max}
				\]
			with $\alpha_b=\frac{v_b-\overline{v}_b}{v_{max}- \overline{v}_b}$. Note $\left<p_b,c_b\right>=v_b$ and $\left<p_s,c_b\right>\geq v_s$ for all $s\in S$. Hence, $c_b$ fully extract the rents from $b$ and doesn't violate any incentive compatibility constraint. 
		\end{enumerate}
	Repeating this process for all other $b'\in \hat{B}$ we obtain a feasible contract menu which achieves full extraction.
\end{proof}

This result shows how full extraction is maintained if we replace the condition over the beliefs of the behavioral types with a condition over their payoffs. While the condition is restrictive, it is less extreme than the condition required for fully strategic types which replaces the inequality in the proposition above with an equality.

\section{Auction application}\label{sec:auction}

In this section we introduce an auction environment to illustrate the main result. We start by formally describing the auction model, then we use a single bidder reduction and apply our main theorem and characterize the fully extracting mechanism.

We consider a standard private values auction environment with correlation: there is a single item which could be allocated to one of $n\geq 2$ bidders. The set of buyers is denoted by $N$. Each bidder has a valuation $\theta_i$ for the item. This valuation is each buyer private information, and hence only known to himself. There are a finite set of potential valuations for each bidder $i$, which we denote by $\Theta_i$. We also define $\bm\Theta=\times_{i\in N} \Theta_i$ and $\Theta_{-i}=\times_{j\neq i} \Theta_j$, with general elements $\bm\theta$ and $\theta_{-i}$ respectively. There is a common prior $F$ over the vector of valuations $\bm\theta$, i.e., $F\in\Delta(\bm\Theta)$. 

We are interested in the case of correlated valuations, so we do not impose any independent distributions assumption over $F$. This implies that a bidder $i$ with valuation $\theta_i$ holds beliefs $F(\cdot|\theta_i)\in \Delta(\Theta_{-i})$ over the valuations of the other bidders. 

As in the general model of Section \ref{sec:model}, we introduce some \emph{behavioral types} among the bidders. Here a behavioral type will be determined by his valuation-belief pair. We denote by $B_i\subseteq \Theta_i$ the set of behavioral types for player $i$. Here behavioral types will always report truthfully while non-behavioral types will report what is best for them. Invoking the revelation principle, we will focus on direct revelation mechanisms without loss.

For the discussion of this section, we focus on the symmetric case in which all bidders share the same space of valuations, that is for any $i$ and $j$, $\Theta_i=\Theta_j=\Theta$, and their beliefs are also symmetric, that is for any $t\in \Theta$ and $\omega\in \Theta^{n-1}$, $F(\theta_{-i}=\omega|\theta_i=t)=F(\theta_{-j}=\omega|\theta_j=t)$ for all $i$ and $j$. Moreover, we assume $B_i=B_j=B$ for all bidders $i$ and $j$ as well.\footnote{This is just for simplifying the exposition, and all the discussion extends directly to the asymmetric case.}

We further assume that each valuation generates a different distribution over the valuations of the other bidders, so $F(\cdot|\theta_i=t)\neq F(\cdot|\theta_i=t')$ for all $t,t'\in\Theta$.

Note that by our symmetry assumption, each valuation will not only determine the beliefs but also the degree of sophistication of a particular bidder of type $\theta$. Hence, if he has valuation $\theta\in B$, then the bidder will hold beliefs $F(\cdot|\theta)$ and always report truthfully, while if his valuation is $\theta'\not\in B$ then his beliefs are $F(\cdot|\theta')$ and he is fully strategic.


We proceed to introduce a single bidder reduction of the auction described above, in which we will focus the analysis of the problem from a perspective of a single bidder taking expectations over the information of the other bidders as required. This reduction will allows us to use the main theorem above to solve for the optimal auction. 

Since we are interested in the question of when full surplus extraction is feasible, we will fix the allocation rule to be the one maximizing the total surplus, i.e., the efficient allocation in which the bidder with the highest valuation gets the item. Moreover, we will assume that any tie is resolved in favor of a particular bidder $i$ and focus on the analysis of this bidder.\footnote{We do this only for simplicity, everything extends directly to any alternative tie-breaking rule as usual.}

We also use the same notation from Section \ref{sec:model} for the elements of the single bidder reduction, so we can use Theorem \ref{thm} directly. 

In particular, we denote valuations of bidder $i$ by $t$ and the vector of valuations of the other bidders different from $i$ by $\omega$. We let the beliefs of type $t$ be $p_t(\omega)=F(\theta_{-i}=\omega|\theta_i)$.

We will denote the gross expected utility of bidder $i$ with valuation $t$ by $v_t$. Hence, under the efficient allocation rule $v_t$ will be equal to the valuation $t$ multiplied by the probability he has the highest valuation.\footnote{Given the notation of this section, we would have \[
		v_t=t\cdot\left(\sum_{\{\theta_{-i}:\max_{j\neq i}\theta_j\leq t\}} F(\theta_{-i}|\theta_i=t)\right).
	\]}

Finally, $c_t(\omega)$ will represent the transfer made by the bidder if his reported valuation is $t$ and the valuations reported by other bidders is $\omega$.

In order to apply Theorem \ref{thm}, we will need to impose some conditions over the beliefs of the bidder. In particular, we impose that for all $t\in \Theta$

	\[
		p_t\not\in co(p_{t'}:t'\not\in B \mbox{ and }t'\neq t).
	\] 

Note that this condition is equivalent to the two conditions in Theorem \ref{thm}. Hence, we can apply directly Theorem \ref{thm} to guarantee full surplus extraction in this setting. Moreover, we can use the construction in the proof of Theorem \ref{thm} to compute the transfers required to achieve it in this setting. Since this transfer rule is incentive compatible and extracts all the surplus under the surplus maximizing allocation, the optimal mechanism will indeed extract all the informational rents in expectation as long as the condition above holds. We state this result formally in the corollary below, using the original notation for the auction environment. Note that we can drop the symmetry assumption without any loss.

	\begin{proposition}
		Consider the auction environment. Let $B_i$ the set of behavioral types for bidder $i$. If for all bidders $i$, and valuations $\theta_i\in \Theta_i$, 
			\[
				F(\cdot|\theta_i) \not\in co(\{F(\cdot|\theta'_i):\theta'_i\not\in B_i \mbox{ and }\theta'_i\neq \theta_i\})
			\] 
			then the optimal mechanism achieves full surplus extraction.
	\end{proposition}

	\begin{proof}
		Follows from Theorem \ref{thm} and the single bidder reduction characterized above.
	\end{proof}

Note that the above analysis will remain essentially unchanged if in addition to the private information of other bidders, there we include other variables correlated with the valuation of bidder $i$ in which the auction could condition the payments (and allocation).

\section{Concluding remarks}\label{sec:end}

We examined the full surplus extraction problem considering the presence of behavioral types who always reveal their information truthfully. We characterize the conditions that guarantee full surplus extraction regardless of the valuations of the agents. The key condition is a weakened version of the one originally identified by \citet{cremermclean1988} in the case without behavioral types. We show that extracting all rents from behavioral types is possible regardless of the payoffs of other types if the beliefs of the behavioral types are not in the convex hull of the beliefs of the strategic types. We also provide a version of full surplus extraction if we replace the condition over the beliefs of behavioral types by a condition on their payoffs. Finally, we applied our main result to a auction environment with correlated valuations.

While the assumption over the behavior of behavioral types in this model seems extreme, it allows to study a simple environment and provide a characterization in a setting where the designer could take full advantage from this type. This serves as an important starting point to formally study the limits of surplus extraction in environments with correlation and non-fully strategic agents. Including richer assumptions on the behavior of non-strategic agents seems like an interesting path to pursue in future research.

\bibliography{behavioralmd}{}

\end{document}